\documentclass[
superscriptaddress,
amsmath,
amssymb,
achemso,
aps,
prl,
twocolumn,
nofootinbib,
]{revtex4-2}

\usepackage{graphics,amssymb,amsmath,epsfig,color}
\usepackage{mathtools} 
\usepackage{graphicx}
\usepackage{setspace}
\usepackage{hyperref}
\usepackage{braket}

\begin{document}

\title{Vibronic coupling limits the use of high-lying electronic states in complex molecules for laser cooling}

\author{Haowen Zhou}
\thanks{These authors contributed equally.}
\affiliation{Department of Physics and Astronomy, University of California, Los Angeles, California 90095, USA}

\author{Paweł Wójcik}
\thanks{These authors contributed equally.}
\affiliation{Department of Chemistry, University of Southern California, Los Angeles, California 90089, USA}
\author{}
\thanks{Present address: Department of Chemistry and Biochemistry, Florida State University, Tallahassee, Florida 32306, USA} 
\noaffiliation

\author{Guo-Zhu Zhu}
\altaffiliation[Present address: ]{Department of Chemistry, Fudan University, Shanghai 200438, China}
\affiliation{Department of Physics and Astronomy, University of California, Los Angeles, California 90095, USA}

\author{Guanming Lao}
\altaffiliation[Present address: ]{Pritzker School of Molecular Engineering, University of Chicago, Chicago, IL 60637}
\affiliation{Department of Physics and Astronomy, University of California, Los Angeles, California 90095, USA}

\author{Taras Khvorost}
\affiliation{Department of Chemistry and Biochemistry, University of California, Los Angeles, California 90095, USA}

\author{Justin R. Caram}
\affiliation{Department of Chemistry and Biochemistry, University of California, Los Angeles, California 90095, USA}

\author{Wesley C. Campbell}
\affiliation{Department of Physics and Astronomy, University of California, Los Angeles, California 90095, USA}
\affiliation{Center for Quantum Science and Engineering, University of California, Los Angeles, California 90095, USA}
\affiliation{Challenge Institute for Quantum Computation, University of California, Los Angeles, California 90095, USA}

\author{Anastassia N. Alexandrova}
\affiliation{Department of Chemistry and Biochemistry, University of California, Los Angeles, California 90095, USA}
\affiliation{Center for Quantum Science and Engineering, University of California, Los Angeles, California 90095, USA}

\author{Anna I. Krylov}
\affiliation{Department of Chemistry, University of Southern California, Los Angeles, California 90089, USA}

\author{Eric R. Hudson}
\affiliation{Department of Physics and Astronomy, University of California, Los Angeles, California 90095, USA}
\affiliation{Center for Quantum Science and Engineering, University of California, Los Angeles, California 90095, USA}
\affiliation{Challenge Institute for Quantum Computation, University of California, Los Angeles, California 90095, USA}

\date{\today}
\bigskip

\begin{abstract}
    Laser cooling of large, complex molecules is a long-standing goal, instrumental for enabling new quantum technology and precision measurements.
    A primary consideration for the feasibility of laser cooling, which determines the efficiency and technical requirements of the process, is the number of excited-state decay pathways leading to vibrational excitations.
    Therefore, the assessment of the laser-cooling potential of a molecule begins with estimate of the vibrational branching ratios of the first few electronic excited states theoretically to find the optimum cooling scheme.
    Such calculations, typically done within the BO and harmonic approximations, have suggested that one leading candidate for large, polyatomic molecule laser cooling, alkaline earth phenoxides, can most efficiently be laser-cooled via the third electronically excited ($\tilde{C}$) state.
    Here, we report the first detailed spectroscopic characterization of the $\tilde{C}$ state in CaOPh and SrOPh.
    We find that non-adiabatic couplings between the $\tilde{A}$, $\tilde{B}$, and $\tilde{C}$ states lead to substantial mixing, giving rise to vibronic states that enable additional decay pathways. 
    Based on the intensity ratio of these extra decay channels, we estimate a non-adiabatic coupling strength of $\sim 0.1$ cm$^{-1}$.
    While this coupling strength is small, the large density of vibrational states available at photonic energy scales in a polyatomic molecule leads to significant mixing.
    Only the lowest excited state $\tilde{A}$ is exempt from this coupling because it is highly separated from the ground state.
    Thus, this result is expected to be general for large molecules, and implies that only the lowest electronic excited state should be considered when judging the suitability of a molecule for laser cooling. 
\end{abstract}
\maketitle

\section{Introduction}
\begin{figure*}
\centering
\includegraphics[width=1\linewidth]{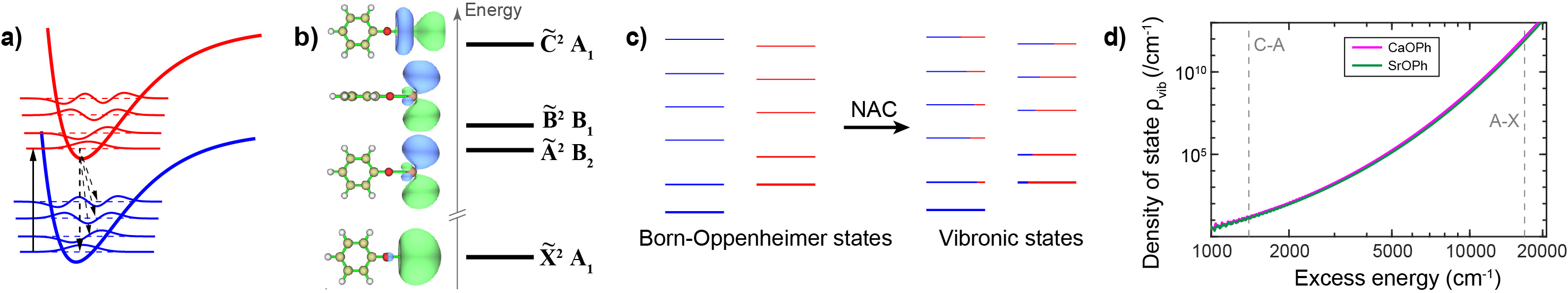}
\caption{
(a) Optical cycling  -- laser transitions (solid lines) and decay pathways (dashed lines). In laser-coolable molecules, the spontaneous emission must be confined only to a few vibronic branchings with a dominant decay to the ground vibrational level. 
(b) Molecular orbitals of the ground $\tilde{X}$ and first three excited electronic states $\tilde{A}$, $\tilde{B}$, and $\tilde{C}$ of the MOPh (M = Ca,Sr) molecules. These metal-centered molecular orbitals ensures similar potential energy surface across these states, leading to favorable FCFs between the same vibrational levels.
(c) The effect of non-adiabatic couplings (NAC) between BO states. 
Red and blue colors denote two (adiabatic) electronic states (lines) and respective vibrational states. 
NAC gives rise to vibronic states that have contributions from both electronic states. 
(d) Density of MOPh (M = Ca,Sr) vibrational states as a function of excess energy within a single electronic state. 
For complex molecules, the higher electronic states resides in the dense vibrational levels of the lower electronic states, enhancing the effect of NACs.
The two dashed lines indicate the relative energy difference between $\tilde{C}-\tilde{A}$, and $\tilde{A}-\tilde{X}$ electronic states. 
\label{fig:FCFs}}
\end{figure*}

State-controlled molecules at ultracold temperatures enable broad applications, including molecular quantum information platforms~\cite{yu_scalable_2019,campbell2020dipole, hudson2021laserless, holland_-demand_2023, demille_quantum_2024, cornish_quantum_2024}, precision search of fundamental physics~\cite{augenbraun_molecular_2020, augenbraun_laser-cooled_2020, Hutzler_polyatomic_2020, anderegg_quantum_2023}, and chemistry at low temperatures~\cite{balakrishnan_perspective_2016,puri_synthesis_2017, hu_direct_2019, gregory_moleculemolecule_2021, chen_ultracold_2024}. 
Extending this control to large molecules can bring more complex organic structures into the quantum regime, allowing chemical, and even biochemical, studies in unprecedented details.
Laser-cooling, where the interaction of a system with photons is engineered to provide dissipation, is one of the leading techniques for state-controlled molecule production~\cite{phillips_nobel_1998, rosa_laser-cooling_2004}. 
Laser-cooling relies on repeated optical cycling --- excitation-emission between two electronic states --- which is straightforward to perform in atoms but more challenging in molecules because of their vibrational degrees of freedom. 
For successful laser-cooling, the emission should  return the molecule primarily to the same vibrational state from which it was excited, creating a complete optical cycle. 
However, it is common for molecules to populate additional vibrational states in the emission process, as illustrated in Fig. \ref{fig:FCFs}a. 
The difficulty of finding molecules in which vibrational branching is suppressed has limited the use of this technique to a few small molecules~\cite{fitch_laser-cooled_2021, shuman_laser_2010, anderegg_radio_2017, collopy_3d_2018, baum_1d_2020, kozyryev_sisyphus_2017, mitra_direct_2020}. 
The effect of undesirable vibrational branching can be mitigated by using additional lasers, but in large molecules this approach soon becomes impractical due to a large number of vibrational states accessible upon decay~\cite{augenbraun_direct_2023}. 
Within the Born--Oppenheimer (BO) framework, the extent of vibrational branching is typically quantified by the Franck--Condon factors (FCFs) --- overlaps between vibrational wavefunctions of the two electronic states (Fig. \ref{fig:FCFs}a). 
In an ideal molecule, the FCFs are `diagonal' -- that is, only transitions between states with the same vibrational quantum numbers are possible. 
In the BO framework this can be achieved when the potential energy surfaces of the two cycling states are parallel -- a situation uncommon in chemistry since the electronic excitation generally affects molecular bonding, leading to significant changes in the shape of the potential energy surface. 

One solution to this problem is to functionalize large molecules with an optical cycling center (OCC)~\cite{kozyryev_proposal_2016, isaev_polyatomic_2016, klos_prospects_2020, zhou2025vibronic}, such as an alkaline earth(I)-oxygen moiety in which the excitation is localized on the metal atom. 
Fig. \ref{fig:FCFs}c illustrates this concept by showing the ground ($\tilde{X}$) and first three excited states ($\tilde{A}$, $\tilde{B}$, and $\tilde{C}$) of OCC-functionalized molecules MOPh (M = Ca,Sr, Ph = phenyl). 
When the M(I)-O moiety is attached to the organic scaffold, optical excitation remains localized and is thus decoupled from the rest of the molecule. 
Hence, the electronic transition causes minimal changes to the molecular structure, so that only a few vibrational channels might be populated in the course of cycling.
This allows favorable optical cycling transitions to be maintained with increasing molecular size~\cite{mitra_direct_2020, zhu_functionalizing_2022}.
In fact, OCC functionalization have been successfully demonstrated for complex molecules such as arenes ~\cite{Maxim:CoolingLarge:20, dickerson_optical_2021, mitra_pathway_2022} and diamondoids ~\cite{lao_bottom-up_2024, guo_surface_2021} with the diagonal vibrational branching ratio (VBR) exceeding 90\%.
These favorable VBRs can be further improved with appropriate chemical modifications of the molecular scaffold~\cite{dickerson_franck-condon_2021}.

In Fig. \ref{fig:FCFs}b, the ground state of MOPh is $s$-like and the three excited states are $p$-like. 
Because of the molecular scaffold, the three-fold degeneracy is lifted, but all three excited states show similarly high oscillator strengths, as expected for $s-p$ type transitions. 
Hence one can, in principle, consider cycling schemes using any of the three states.
Calculations of FCFs show better (more diagonal) VBRs from the higher electronic state, $\tilde{C}$, than from the two lower states, which can be rationalized by a more localized orbital. 
The computed VBR for the $\tilde{C}\rightarrow\tilde{X}$ transition is $\sim99\%$, which is a significant improvement compared to the $\sim96\%$ VBRs for the decay from the $\tilde{A}$ and $\tilde{B}$ states. 
This highly favorable VBR has motivated experimental examination of the $\tilde{C}$ state.

For diatomic and triatomic molecules such as CaH and CaOH, it is possible to use higher electronic states for optical cycling~\cite{vilas_magneto-optical_2022, vazquez-carson_direct_2022}.
However, it is an open question if higher electronic states can be leveraged for optical cycling in larger molecules.
In most photochemical processes in collisional environments, non-radiative relaxation from the higher electronic states to the lowest excited state is so effective that only fluorescence from the lowest excited state is usually observed ; this is known as Kasha's rule\cite{kasha_characterization_1950}. 
However in collision-free environments, as is typical for laser cooling, conservation of energy dictates Kasha's rule does not apply and the primary concern are non-adiabatic effects,  
given the dramatically increased density of rovibronic states. 
These non-adiabatic couplings (NAC) result in the breakdown of the BO (or adiabatic) approximation, which can dramatically alter the predicted VBRs.
Though in previous work, we have explored how anharmonic couplings within a single electronic state enable Fermi resonances between a strong vibrational fundamental mode and weak vibrational combination mode and alter the VBRs via intensity borrowing~\cite{zhu_extending_2024},
it was not known if the localized nature of the OCC transition provided enough suppression of the NACs to allow laser cooling via higher lying electronic states.

Fig. \ref{fig:FCFs}c illustrates the consequences of these NACs. 
In the BO framework, the vibrational states of the two electronic manifolds are non-interacting and the probability of transitions are determined by products of the electronic transition dipole moments and vibrational overlaps (FCFs). 
The NACs cause state mixing and the resulting vibronic states have contributions from more than one electronic state. 
In the context of laser cooling, these mixed vibronic states lead to additional decay channels.
Given the dense manifolds of vibrational states in polyatomic molecules (see e.g. Fig.~\ref{fig:FCFs}(d)), even small non-adiabatic interactions might lead to significant mixing and the resulting additional decays might out-compete the main optical cycling transition.
Therefore, these couplings must be carefully examined when designing laser cooling schemes involving higher electronic states~\cite{ivanov_vibronic_2016}.

In this work, we focus on the optical properties of the $\tilde{C}$ state of two OCC-functionalized molecules MOPh (M = Ca,Sr, see Fig. \ref{fig:FCFs}c), and study their spectroscopy in detail.
In both cases, the $\tilde{C}$ state shows signatures of NACs to rovibronic states of the lower $\tilde{A}$ and $\tilde{B}$ states. 
In what follows, we present experimental results showing that NACs result in additional decay pathways, with their intensities similar or even stronger to the main optical cycling transition.
To describe these additional features, the theory must go beyond the BO approximation.
Here we employ a vibronic Hamiltonian approach of K{\"o}ppel, Domcke, and Cederbaum -- the KDC Hamiltonian\cite{Cederbaum:LVC:84,KDC:81,Koppel:CIbookCh7:04}, which enables calculation of vibronic states and transitions between them.
When combined with high-accuracy quantum-chemistry methods, such as equation-of-motion coupled-cluster theory\cite{Stanton:93:EOMCC, Nooijen:EOMEA:95, Bartlett:CC_review:07,Krylov:EOMRev:07,Krylov:OSRev}, the KDC approach was shown to accurately describe complicated vibronic spectra in various systems\cite{Koppel:02,Stanton:NO3:07,wojcik2024vibronic}.
We present results of such calculations, which provide insight into the observed spectra.
As a result, we conclude that \emph{large polyatomic molecule laser cooling schemes should only consider use of the lowest electronic state}. 

\section{Results and discussion}

\begin{figure*}
    \centering
    \includegraphics[width=1\linewidth]{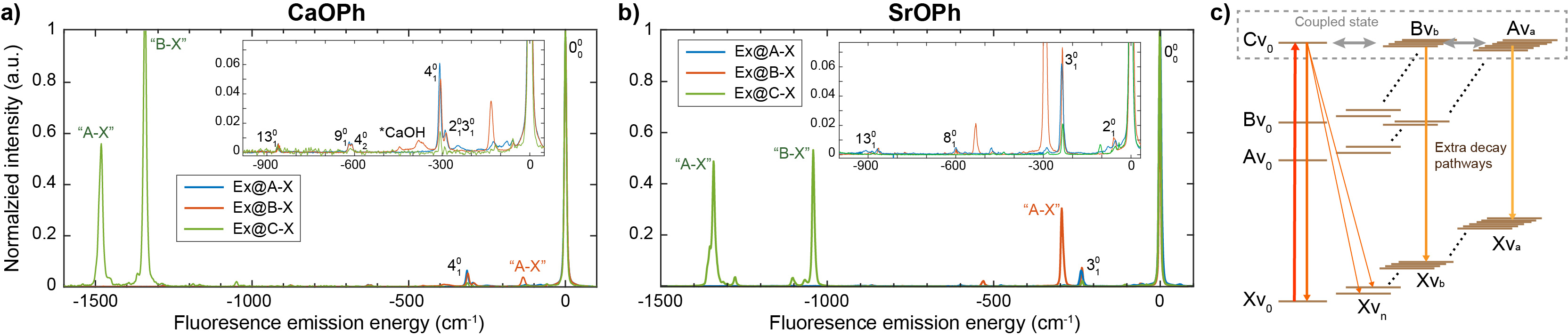}
    \caption{(a,b) The DLIF spectra of CaOPh and SrOPh from the three excited states to the ground state. The spectra are normalized and centered with respected to the 0-0 transition to facilitate the comparison. In the insets we zoom in for branching decays into the first few vibrational modes. Peaks are labeled as $\nu_{i}^{f}$, where $\nu$ denotes the vibrational mode, $i$ and $f$ define the vibrational quantum number in the excited and ground electronic state. The labels "Ex@A/B/C-X" indicate that the molecule is excited from the ground $\tilde{X}$ state into the vibrationless level of the excited $\tilde{A}/\tilde{B}/\tilde{C}$ state respectively, and the subsequent fluorescence spectra are taken. (c) The schematic of fluorescence decays from the coupled $\tilde{C}$ state as in Eq.~\ref{eq:coupling}. The ``A-X'' and ``B-X'' DLIF peaks are the vertical decays from the coupled $\tilde{A}$ and $\tilde{B}$ components.}
    \label{fig:DLIF}
\end{figure*}

Fig.~\ref{fig:DLIF}(a,b) shows the measured dispersed laser-induced fluorescence (DLIF) spectra obtained by exciting the molecules into the $v_0$ levels of the three electronic states, and recording the subsequent fluorescence as a function of emission wavelength.
These spectra are normalized to facilitate comparison; each peak represents a decay channel to a vibrational level of the ground state.
Similar to the previously reported $\tilde{A}$ and $\tilde{B}$ states DLIF spectra~\cite{zhu_functionalizing_2022, lao_laser_2022}, the decay channels of the $\tilde{C}$ state occur at the same relative frequencies, confirming that this state is localized on the metal atom and that the molecular structure changes minimally upon $\tilde{C}\rightarrow\tilde{X}$ transition.
The DLIF of the $\tilde{C}$ shows, as expected from theory, that vibration changing decay to the lower frequency modes is indeed suppressed relative to decays from the $\tilde{A}$ and $\tilde{B}$ states. 
(See the inset of Fig.~\ref{fig:DLIF}. A table of measured and computed VBRs of several MO-stretch modes for the three electronic states is given in the Supplementary Information.)
However, the $\tilde{C}$ state exhibits strong decays at frequencies associated with diagonal decays from $\tilde{A}\rightarrow \tilde{X}$ and $\tilde{B}\rightarrow \tilde{X}$, which are labeled in Fig.~\ref{fig:DLIF} as ``$\tilde{A}-\tilde{X}$" and ``$\tilde{B}-\tilde{X}$".
With intensities similar or even stronger than the diagonal decay on the $\tilde{C}\rightarrow \tilde{X}$ transition, these decays       significantly complicate optical cycling via the $\tilde{C}$ state.

At first glance, these peaks appear to arise from the $\ket{\tilde{A},\nu_0}$ and $\ket{\tilde{B},\nu_0}$ states, which could be produced by collisional relaxation of molecules excited to the $\tilde{C}$ state. 
However, careful assessment of their relative strengths as a function of molecule density shows no variation (see SI). 
Therefore, we conclude that these decays result from intrinsic properties of the $\tilde{C}$ state. 
A potential origin of these peaks is from the diagonal relaxation of  \emph{vibrationally    excited}  $\tilde{A}$ and $\tilde{B}$ states that are non-adiabatically coupled to the $\ket{\tilde{C},\nu_0}$ state.
Since the $\tilde{C}$ state is higher in energy ($\sim1400$ cm $^{-1}$ above the $\tilde{A}$ state), it is embedded in a dense rovibronic manifold of the lower electronic states. 
Fig.~\ref{fig:FCFs}d shows the density of vibrational states of CaOPh and SrOPh calculated as a function of excess energy.
As can be seen, the density of the excited vibrational levels of the lower $\tilde{A}$ and $\tilde{B}$ states reaches the order of $\sim10$~/cm~$^{-1}$ near the origin of the $\tilde{C}$ state.
As such, the excited molecular state is no longer a pure $\tilde{C}$ state, but has significant contributions from other vibrational levels originating from the $\tilde{A}$ and $\tilde{B}$ manifolds, i.e.:
\begin{equation}
\label{eq:coupling}
\ket{\Psi} = c_{{\nu_0}}\ket{\tilde{C},\nu_0} + \sum_{\nu_a}c_{\nu_a}\ket{\tilde{A},\nu_a} + \sum_{\nu_b}c_{\nu_b}\ket{\tilde{B},\nu_b}, 
\end{equation}
where $\nu_a$/$\nu_b$ denote vibrational states from the respective manifolds and $c_{\nu_A}/c_{\nu_B}$ denote the mixing coefficient of each state. 
In this scenario, the ``$\tilde{A}-\tilde{X}$'' and ``$\tilde{B}-\tilde{X}$'' relaxation features result from decay of these mixed $\ket{\tilde{A},\nu_a}$ and $\ket{\tilde{B},\nu_b}$ components into $\ket{\tilde{X},\nu_a}$ and $\ket{\tilde{X},\nu_b}$ states.
For an OCC-functionalized molecule, these diagonal relaxations are the most probable, and happen at similar wavelength as the $\ket{\tilde{A},\nu_0} \rightarrow \ket{\tilde{X},\nu_0}$ and $\ket{\tilde{B},\nu_0} \rightarrow \ket{\tilde{X},\nu_0}$ decays ~\cite{zhu_extending_2024}.
The strong intensity of these features indicates substantial coupling.

\begin{figure*}
    \centering
    \includegraphics[width=0.9\linewidth]{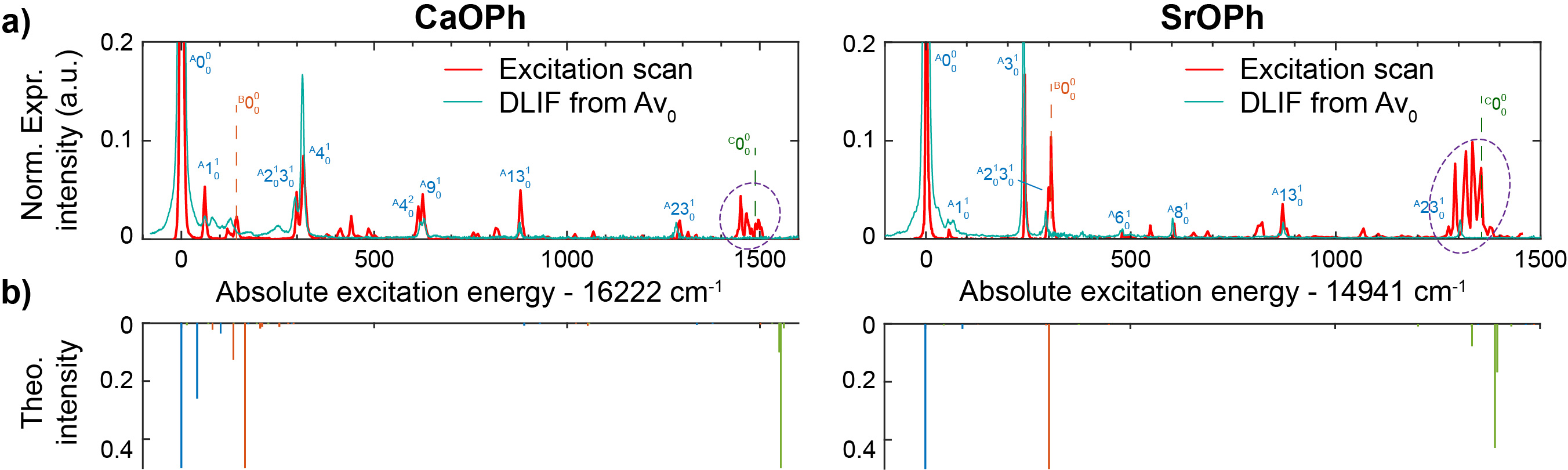}
    \caption{(a) Measured excitation spectra (red line) for CaOPh (left) and SrOPh (right) probing at ``$\tilde{A}-\tilde{X}$'' diagonal relaxations. The DLIF scan (light cyan) from $\ket{\tilde{A},\nu_0}$ state is shown for comparison. Near the $\ket{\tilde{C},\nu_0}$ energy, we observe multiple extra excitation peaks which are absent in the DLIF measurements (circled for clear identification).
    (b) Calculated excitation spectra including NAC terms. The colors indicates components from $\tilde{A}$ (blue), $\tilde{B}$ (orange) and $\tilde{C}$ (green) states. Because it is not practical to include NACs from all vibronic states, only a few vibrational modes are included in the calculation. The qualitative match of the multiple peak structures near the $\tilde{C}$ state confirms the presence of NACs.}
    \label{fig:excitation}
\end{figure*}

To further probe the vibronically mixed character of the $\tilde{C}$ state, we recorded the excitation spectra by scanning the excitation laser frequency while monitoring the fluorescence detection at the aforementioned vertical relaxation energies.
Here, the excitation process, e.g. $\ket{\tilde{A},\nu_a} \leftarrow \ket{\tilde{X},\nu_0}$, leads to peak intensities that correlate with the FCFs, which parallel with the intensities in the corresponding relaxation process $\ket{\tilde{A},\nu_0} \leftarrow \ket{\tilde{X},\nu_a}$ in the DLIF spectra. 
However, due to saturation effects from the large available laser power more vibration changing transitions can be resolved~\cite{zhu_extending_2024}. 
Fig.~\ref{fig:excitation}a shows the measured excitation spectra while probing at the ``$\tilde{A}-\tilde{X}$'' relaxation wavelength.
At low vibrational energies, most peaks can be assigned based on the same vibrational branchings in the DLIF scan (the pale cyan line in  Fig.~\ref{fig:excitation}a).
However, near the $\tilde{C}-\tilde{X}$ 0-0 transition energy, many new features are observed exclusively in the excitation spectra.
Their absence in the DLIF spectra indicates that the corresponding vibrational modes have negligible FCFs.
As such, the presence of these peaks must result from NACs, which create vibronically-mixed states with the energetically close $\ket{C,\nu_0}$ level.

This interpretation of the observed spectra are supported by calculations using the KDC vibronic Hamiltonian.
Here we consider linear vibronic coupling as a leading electronic 
effect, while noting that other types of couplings are also possible, e.g., spin-orbit, spin-rotation, etc.~\cite{lefebvre2004spectra}, we only consider linear vibronic coupling as a leading effect.
Fig.~\ref{fig:excitation}(b) shows the calculated excitation spectra for comparison.
Near the $\ket{\tilde{C},\nu_0}$ state the multiple peak structure is roughly reproduced in the theoretical spectra, even with only linear vibronic counpling, confirming the presence of extra vibronic levels; a perfect match with experimental data should not be expected as the calculations required to include all nearby vibrational states and various possible NACs are too expensive.
Instead, candidate $\tilde{A}$ and $\tilde{B}$ vibrational modes are selected based on the coupling strengths or proximity of their frequencies (see SI for more details).
Nonetheless, the qualitative match establishes the importance of NACs in understanding such spectra.

A mean coupling strength between $\tilde{C}$ and $\tilde{A}/\tilde{B}$ states can be estimated from the intensity ratios in the DLIF spectra. 
The coefficient of the $\tilde{C}$ state component $c_C$ in Eq.\eqref{eq:coupling} can be estimated from the peak intensities as
\begin{equation}
    \frac{I_C}{I_{total}} = \frac{|c_{\nu_0}|^2 A_C}{\sum_{\nu_i}{|c_{\nu_i}|^2A_{\nu_i}}} = \frac{|c_{\nu_0}|^2 \mu_C^2 \omega_C^3}{\sum_{i,\nu_i}{|c_{i\nu_i}|^2\mu_i^2 \omega_i^3}} \approx \frac{|c_C|^2}{\sum_{i,\nu_i}{|c_{i\nu_i}|^2}} 
    \label{eq:intensity}
\end{equation}
where $\nu_i$ denotes the vibronic state, $A$ is the Einstein coefficient, $\mu$ is the transition dipole moment and $\omega$ is the transition frequency to the ground state.
Here, we approximate that transition dipole moments, as well as the transition frequencies, as equal, given that the main decay channels are diagonal involving no change in vibrations.
For SrOPh, the $\tilde{C},\nu_0$ relaxation peaks represents roughly 37\% of the total intensity in the DLIF spectra.
Given the $\tilde{C}$ state is located $\sim1400$~cm~$^{-1}$ above the $\tilde{A}$ state, the corresponding density of vibrational states is calculated to be $\approx 15$/cm$^{-1}$ (see Fig. \ref{fig:FCFs}d).
Therefore, we estimate from the experimental data that the $\tilde{C}-\tilde{A}$ coupling strength is on the order of $\sim0.1$~cm$^{-1}$.
Similarly, for CaOPh with a slightly higher density of states ($\approx 30$/cm$^{-1}$), the $\tilde{C}$ state intensity ratio is measured to be 0.22, producing a coupling strength of similar scale (see SI for more details.)

These results confirm that NACs between different electronics states results in additional vibration-changing decay pathways that would require repumping for successful optical cycling; this adds more complexity to the already-complicated molecular laser cooling schemes.
While it may be possible to mitigate the effect by shifting the energy levels, for complex molecules with the ever-increasing number of coupled vibrational states avoiding these coupling appears practically impossible.
Since, for a relatively low density of states ($\sim 10/\text{cm}^{-1}$), even a small coupling strength on the order of $0.1 \text{ cm}^{-1}$ can mix a nearby $\tilde{A}/\tilde{B}$ vibrational level into the $\left |\tilde{C},v_0\right \rangle$ state.
As a result, \emph{optical cycling using the $\tilde{A}$ state appears to be the only route for laser cooling large polyatomic molecules.} 

Thus, an interesting question is: can non-adiabatic coupling of the $\ket{\tilde{A},0}$ state to high-lying vibrational states $\ket{\tilde{X},\nu_X}$ be significant?
If so, at sufficient mixing vibration changing decays at infrared wavelengths (e.g. to $\ket{\tilde{X},\nu_X-1}$) could occur and make optical cycling virtually impossible. 
To analyze this question, consider laser excitation in the BO basis to $\ket{\tilde{A},0}$. 
An NAC of strength $\Omega$ leads to population of rovibrational states $\ket{\tilde{X},\nu_x}$ at a rate of $\Gamma_X \approx (2\pi/\hbar) \Omega^2 \rho$, where $\rho$ is the density of $\ket{\tilde{X},\nu_x}$ near $\ket{\tilde{A},0}$.
Assuming this population would not return to $\ket{\tilde{A},0}$, as would be the case for a large density of states and finite lifetime of $\ket{\tilde{X},\nu_x}$ due to radiative decay to other $\ket{\tilde{X},\nu_x'}$ states, this coupling leads to loss from the optical cycling process.
Good optical cycling requires scattering of order $10^4$ photons, thus for a radiative lifetime $\tau$ of the $\ket{\tilde{A},0}$, we require $\tau \Gamma_{X}\leq 10^{-4}$.
As a result, for a typical radiative lifetime of 10~ns, we require $\Omega^2\rho \leq 10^{-8}$~cm$^{-1}$.

The NAC coupling can be estimated 
as $\Omega = \Omega_o F$, where $\Omega_o$ characterizes the strength of the interaction and $F$ is the Franck-Condon factor between $\ket{\tilde{A},0}$ and $\ket{\tilde{X},\nu_X}$.
By construction, OCC molecules have nearly diagonal Franck-Condon factors, thus $F$ can be expected to be extremely small. 
Take SrOPh for example, the $\ket{\tilde{X},\nu_X}$ states near the $\ket{\tilde{A},0}$ state, have an average of approximately 45 vibrational quanta.
As discussed in the SI, a vibrational mode with this many quanta typically has a Franck-Condon factor with the ground state of roughly $F \lesssim 10^{-45}$; this is likely a overestimate since the majority of the combination modes involve vibrations far away from the metal center which have weaker FCFs.
Therefore, optical cycling requires $\rho < 10^{82}/\Omega_o^2$ /cm$^{-1}$.
Thus, even if $\Omega_o$ was as large as the energy of the $\ket{\tilde{A},0}$ (15,000 cm$^{-1}$), the requirement is easily satisfied.
Further, as more atoms are added to the molecule, the number of low-frequency modes increases leading to even smaller estimates for $F$ (See SI Table S4).
Therefore, in general, laser cooling of OCC-functionalized complex molecules does not appear to be limited by NAC to the $\tilde{X}$ state.
This conclusion is supported by recent work that has observed strong \emph{optical} fluorescence in OCC functionalized diamonoids~\cite{lao_bottom-up_2024} with up to 46 atoms.

\section{Conclusion}

In summary, we have studied CaOPh and SrOPh as model systems to examine the NACs and resulting vibronic mixing of the three lowest excited states $\tilde{A}$, $\tilde{B}$, and $\tilde{C}$.
Such couplings mix the $\tilde{C}$ state with the dense vibrational levels of the lower $\tilde{A}/\tilde{B}$ states, giving rise to vibronic states that can decay via additional channels, thus  deteriorating optical cycling efficiency.
Based on the observed mixing, we estimate a coupling strength between the $\tilde{C}$ and $\tilde{A}/\tilde{B}$ states on the order of $\sim0.1$ cm $^{-1}$. 
We note that the observed vibronic coupling is not restricted to the molecules discussed above. 
Similar features show up in the $\tilde{C}$ state spectra of many other OCC-functionalized molecules such as Ca/SrOPh-4-Cl, Ca/SrOPh-4-Br, and Ca/SrOPh-3,4,5-3F (see SI).

The ubiquitous presence of the NACs imposes an intrinsic limit that precludes the use of any high-lying excited electronic states for laser cooling of complex molecules.
As a result, when working with large polyatomic molecules, which exhibit a high density of rovibronic states, laser cooling should be considered only using the first excited state for for efficient optical cycling. 

\section*{Data Availability}
The data that support the findings of this study are openly available \cite{zhou_2025_16897009}.

\section*{Acknowledgments}
We thank David Nesbitt for helpful conversations. AIK and PW are grateful to the late Professor John F. Stanton for his guidance in mastering KDC simulations. 
This work was supported by the NSF Center for Chemical Innovation Phase I (Grant No. CHE-2221453) and NSF PHY-2110421. WCC and ERH acknowledge institutional support from NSF OMA-2016245. 

\bibliographystyle{NoURL}
\bibliography{Cstate,allrefs,pawels_library}

\newpage
\widetext
\begin{center}
\textbf{\large Supplemental Information for: Vibronic coupling limits the use of high-lying electronic states in complex molecules for laser cooling}
\end{center}
\setcounter{equation}{0}
\setcounter{figure}{0}
\setcounter{table}{0}
\setcounter{page}{1}
\setcounter{section}{0}
\makeatletter
\renewcommand{\theequation}{S\arabic{equation}}
\renewcommand{\thetable}{S\arabic{table}}
\renewcommand{\thefigure}{S\arabic{figure}}
\renewcommand{\bibnumfmt}[1]{[#1]}
\renewcommand{\citenumfont}[1]{#1}

\section{Experimental Methods}

\begin{figure}[h!]
    \centering
    \includegraphics[width=0.75\linewidth]{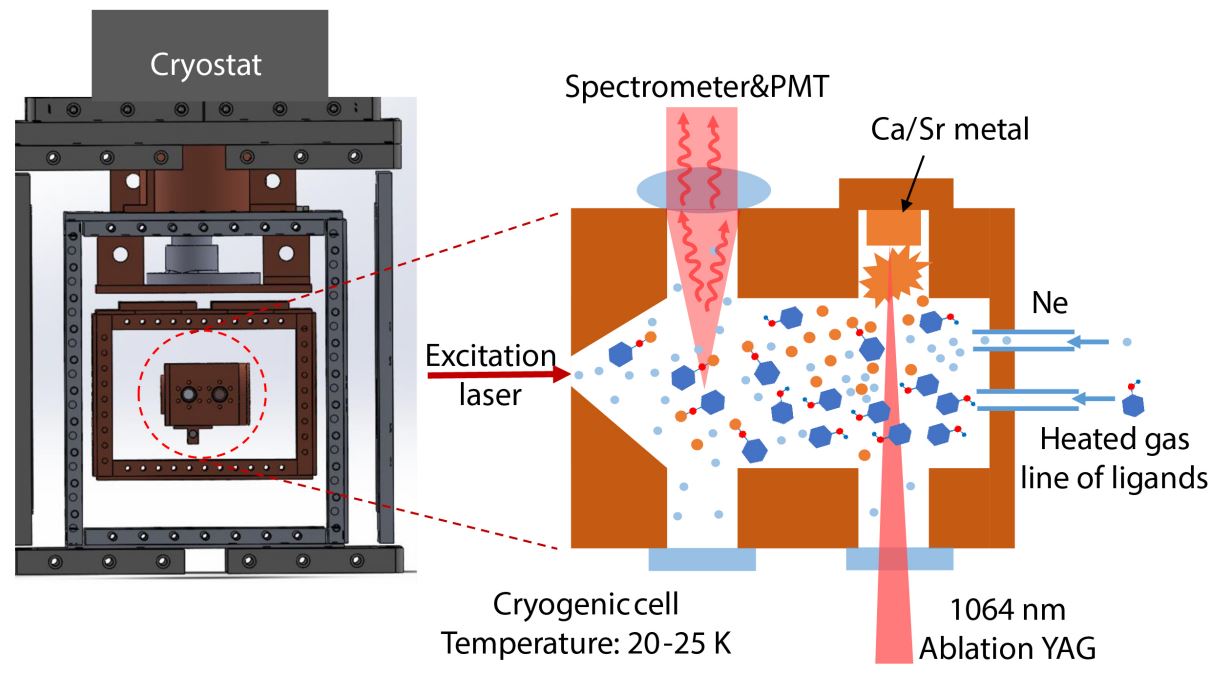}
    \caption{Schematic illustration of the experimental setup.}
    \label{fig:S1}
\end{figure}

The experiments were conducted within a cryogenic buffer gas cell operated at temperatures of $\sim$20 K. CaOPh and SrOPh molecules were produced by reacting the corresponding organic vapor precursor (phenol) with the laser-ablated Ca/Sr metal atoms. All chemicals were purchased from Sigma-Aldrich and used without further purification. The laser ablation was achieved by an Nd:YAG laser (Minilite II) operating at 1064 nm with a pulse energy of approximately 6 mJ and a repetition rate of 10 Hz. The focused spot of the ablation laser was continuously swept over the target using a moving mirror to prevent fluctuations of production yield. The phenol vapor was introduced via a heated gas line from a heated reservoir. The reaction products were subsequently cooled to their vibrational ground states through collisions with Ne buffer gas. We estimate the density of the products to be approximately $10^{15} \text{ cm}^{-3}$. Fig.~\ref{fig:S1} shows the schematic of experimental apparatus.

A tunable pulsed dye laser (LiopStar-E, 10 Hz, linewidth 0.04 $\text{cm}^{-1}$) was used to excite the molecules. Electronically excited molecules then underwent spontaneous emission, resulting in the emission of fluorescence, which was collected by a lens system and directed into a monochromator (McPherson 2035) equipped with a 1200 lines/mm grating. The fluorescence was then detected and imaged by an ICCD camera (Andor iStar 320T). In the experiment, both the excitation and fluorescence detection wavelengths wer e tuned, essentially measuring a full 2D spectroscopy. Specifically, we focused on two types of spectroscopy. In the dispersed laser-induced fluorescence (DLIF) spectroscopy, we fixed the excitation laser at vertical 0-0 transitions from the ground state to the vibrationless level ($\nu_0$) of electronically excited states, and fluoresence detection wavelength was scanned to measure the decays from the excited level. In the excitation spectroscopy, we scanned the excitation wavelength to look for vibrationally excited levels ($\nu_n$) of excited electronic states, and the fluoresence detection was fixed at the vertical relaxation to the same vibrational mode of the ground state. On average, 200-500 shots was taken at each point to ensure reliable signal acquisition.

\newpage
\section{Theoretical Methods}

We carried out vibronic simulations are treated using the KDC Hamiltonian.~\cite{Cederbaum:LVC:84, KDC:81, Koppel:CIbookCh7:04}
The KDC Hamiltonian is set up for use in a basis of diabatic states which for which we use the definition of Ichino, Gauss, and Stanton.~\cite{Stanton:EOMIPdeg:09} 
The KDC Hamiltonian is expanded in the dimensionless normal coordinates of the ground state:
\begin{equation}
    H _{KDC} = 
    H _0 \mathbf{1}
    +
    \begin{bmatrix}
    V ^{(\tilde{A})} & V ^{(\tilde{A}\tilde{B})} & V ^{(\tilde{A}\tilde{C})} \\
    V ^{(\tilde{B}\tilde{A})} & V ^{(\tilde{B})} & V ^{(\tilde{B}\tilde{C})} \\
    V ^{(\tilde{C}\tilde{A})} & V ^{(\tilde{C}\tilde{B})} & V ^{(\tilde{C})} \\
    \end{bmatrix},
    \label{eq:kdc}
\end{equation}
consists of the diagonal harmonic terms
\begin{equation}
    H _0 = -\frac{1}{2} \sum _{i = 1} ^ {N _{nm}} 
    \omega _i 
    \left(
    \partial ^2 _{Q _i}
    + 
    Q _i ^ 2
    \right),
    \label{eq:kdc_H0}
\end{equation}
the diagonal diabatic potential terms
\begin{equation}
    V ^{(\alpha)} 
    = 
    E ^{(\alpha)}
    +
    \sum _{i = 1} ^ {N _{nm}} 
    \kappa _i ^{(\alpha)}
    Q _i
    \label{eq:kdc_Vdiag}
\end{equation}
and the off-diagonal, diabatic coupling terms
\begin{equation}
    V ^{(\alpha\beta)} 
    = 
    \smashoperator{\sum _{i \in \{\text{coupling modes}\}}}
    \lambda _i ^{(\alpha \beta)}
    Q _i
    \label{eq:kdc_Voff}
\end{equation}
In the above, $N_{nm}$ is the number of normal modes included in the model; $\mathbf{1}$ is a $3 \times 3$ identity matrix; $\omega _i$ is the ground state harmonic frequency of the normal mode $i$; $Q_i$ is a dimensionless normal coordinate; $E ^{(\alpha)}$ is the vertical excitation energy of the electronic state $\alpha$;   $\kappa ^{(\alpha)} _i$ is energy gradient of state $\alpha$ along the dimensionless normal mode $i$; and $\lambda ^{(\alpha \beta)} _i$ is the linear diabatic coupling constant between states $\alpha$ and $\beta$ along normal mode $i$.

The electronic structure calculations use the coupled-cluster (CC) based methods.~\cite{Bartlett:CC_review:07}
We computed the vertical excitation energies $E ^{(\alpha)}$ using composite schemes involving basis set extrapolation and use of EOM-CC truncated to include full single, double, and perturbative triple excitations (EOM-CCSD$^*$), where the highest level of theory was especially needed for the $\tilde{C}$ state.~\cite{StantonGauss:SD3:96}
All CC calculations were carried out in \textsc{CFOUR} and \textsc{Q-Chem}.~\cite{cfour, cfour:2020, qchem_feature, qchem5_full}
The parameters of the KDC Hamiltonian, Eq.~\eqref{eq:kdc}, are expanded around the optimized geometry of the $\tilde{X}$ state.
Using definition of (quasi)-diabatic states given by Ichino, Gauss, and Stanton, the linear diabatic coupling constants, $\lambda ^{(\alpha \beta)}_i$, were computed \emph{ab initio}.~\cite{Stanton:EOMIPdeg:09}
We simulated the vibronic spectrum with the \textsc{xsim} program.~\cite{Sharma:xsim_socjt:2024}
The vibronic simulations used 15 basis set functions per vibrational mode and 2000 Lanczos iterations. 
We note that the methods used in this work were recently successful in simulations of vibronic effects in other molecules, in particular for YbOH, CaOH, SrOH, RaOH, SrOCH$_3$, SrNH$_2$, NO$_3$, benzene cation, O$_3$, and pyrazine.~\cite{Doyle:YbOH:20, zhangAccuratePredictionMeasurement2021,
Doyle:SrOH:22, zhangIntensityborrowingMechanismsPertinent2023,
frenettVibrationalBranchingFractions2024, Stanton:NO3:07, Koppel:02,
wojcik2024vibronic}

\newpage
\section{Experimental and theoretical determination of the properties of $\tilde{A}$, $\tilde{B}$ and $\tilde{C}$ states}

\begin{table}[h!]
\centering
\caption{The experimentally and theoretically determined properties of the ground $\tilde{X}$ and first three electronically excited states $\tilde{A}$, $\tilde{B}$ and $\tilde{C}$ of CaOPh and SrOPh}
\begin{tabular}{l|llll|llll}
\hline
\multicolumn{1}{c|}{} & \multicolumn{4}{c|}{CaOPh}            & \multicolumn{4}{c}{SrOPh}     \\ \hline
\begin{tabular}[c]{@{}l@{}} Electronic\\state\end{tabular}  & \begin{tabular}[c]{@{}l@{}} $E_{excitation}$\\(theo.)$^a$\end{tabular} & \begin{tabular}[c]{@{}l@{}} $E_{excitation}$\\(exp.)\end{tabular}  & \begin{tabular}[c]{@{}l@{}} VBR (0-0)\\(theo.)$^b$\end{tabular}  & \begin{tabular}[c]{@{}l@{}} VBR ($\nu_4$)$^c$ \\(exp.)\end{tabular}  & \begin{tabular}[c]{@{}l@{}} $E_{excitation}$\\(theo.)$^a$\end{tabular} & \begin{tabular}[c]{@{}l@{}} $E_{excitation}$\\(exp.)\end{tabular}  & \begin{tabular}[c]{@{}l@{}} VBR (0-0)\\(theo.)$^b$\end{tabular}  & \begin{tabular}[c]{@{}l@{}} VBR ($\nu_3$)$^c$\\(exp.)\end{tabular}
\\  \hline
$X$~$^2A_1$                 & 0                                                                & 0                                                                & -                                                                 & -                                                                   & 0                                                                & 0                                                                & -                                                                 & -                                                                   \\
$A$~$^2B_2$                 & 16491                                                 & 16220                                                            & 0.957                                                             & 0.047                                                               & 13930                                                                  & 14946                                                            & 0.931                                                              & 0.059                                                               \\
$B$~$^2B_1$                 & 16647                                                   & 16357                                                            & 0.973                                                             & 0.042                                                               & 14241                                                                 & 15252                                                            & 0.949                                                              & 0.054                                                               \\
$C$~$^2A_1$                 & 19233                                                                 & 17711                                                            & 0.991                                                               & 0.024                                                               & 15876                                                                 & 16292                                                            & 0.991                                                              & 0.031                                                               \\ \hline
  \multicolumn{9}{l}{\small{$^a$ EOM-EA-CCSD with SOC included.}}\\
  \multicolumn{9}{l}{\small{$^b$ DFT with PBE0/def2-tzvppd.}}\\
  \multicolumn{9}{l}{\small{$^c$ Corresponding to the metal-oxygen stretching vibrational mode.}}\\
\end{tabular}
\end{table}

\section{Vibrational modes included in the excitation spectrum calculations}

\begin{table}[h!]
\centering
\caption{Vibrational modes used in the theoretical computation of excitation spectra}
\begin{tabular}{ll|ll}
\multicolumn{2}{c|}{CaOPh}    & \multicolumn{2}{c}{SrOPh}   \\
\hline
Vibrational mode $   $ & Frequency (cm$^{-1}$) & Vibrational mode $   $ & Frequency (cm$^{-1}$)\\ \hline
$\nu_1$   &  66  &   $\nu_1$   &  65   \\
$\nu_2$   &  77  &   $\nu_2$   &  84   \\
$\nu_{13}$   &  889  &   $\nu_{14}$   &  903   \\
$\nu_{23}$   &  1324  &   $\nu_{23}$   &  1341   \\
$\nu_{24}$   &  1337  &   $\nu_{24}$   &  1349    \\ \hline
\end{tabular}
\end{table}

\newpage
\section{Determination of the $\tilde{C}-\tilde{A}$ coupling strength}

\begin{figure}[h!]
    \centering
    \includegraphics[width=0.7\linewidth]{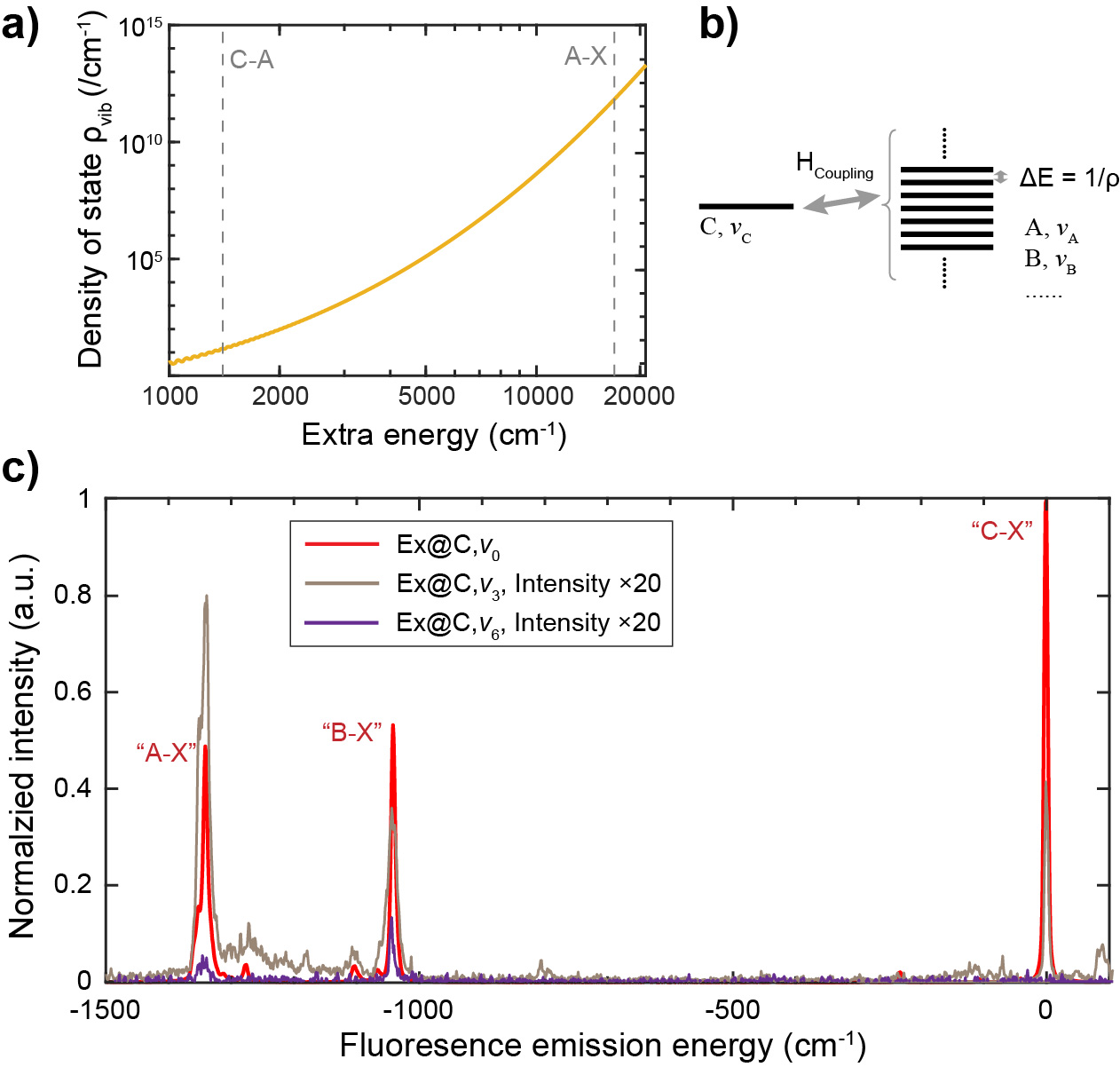}
    \caption{(a) Calculated density of states as a function of excess energy for SrOPh. Two dashed lines indicate the relative energy differences for $\tilde{C}-\tilde{A}$ and $\tilde{A}-\tilde{X}$. (b) A schematics of the model used to calculate the coupling strength. The coupled states are treated as a number of evenly-spaced states with a constant mean coupling strength to $C,\nu_C$. The coupling strength $H_{coupling}$ is calculated by solving the eigenstate of such a coupling system. (c) Comparison between the measured DLIF spectrum for three $\tilde{C}$ state vibrational excitation levels, $\tilde{C},\nu_0$, $\tilde{C},\nu_3$, and $\tilde{C},\nu_6$ of SrOPh. The spectra are normalized to the highest peak in order to facilitate comparison. The intensity of the $\tilde{C}$ state peak drops dramatically as we increase the excitation energy.}
    \label{fig:Cvx}
\end{figure}

We estimate the $\tilde{C}-\tilde{A}$ coupling strength by measuring the $\tilde{C}$ state coefficient $c_C$ in the coupled state (Eq. \ref{eq:coupling},\ref{eq:intensity}). The coefficient $c_C$ is then used to determine the coupling strength. The procedure starts by computing the density of vibrational levels of the lower $\tilde{A}$ state at the corresponding energy. Fig.~\ref{fig:Cvx}(a) shows the density of vibrational states of CaOPh and SrOPh calculated as a function of excess energy using MultiWell program.\cite{barker_multiwell_2023, barker_multiplewell_2001, barker_energy_2009} At $\sim1400$ cm $^{-1}$ excess energy, the density of the excited vibrational levels of the lower $\tilde{A}$ and $\tilde{B}$ states reaches the order of $\sim10$ /cm $^{-1}$. Therefore, it is better to use a mean coupling strength to describe the system instead of focusing on a few specific states. Here, we build our model by coupling the state of interest ($\tilde{C}$ state) with a set of evenly-spaced states (separated by $\Delta E$ determined by the density of states) using a mean coupling strength (See Fig. \ref{fig:Cvx}(b)) By solving the eigenstate of such a system, the mean coupling strength can be determined knowing the $\Delta E$ as well as the $\tilde{C}$ state coefficient $c_C$.

Fig. \ref{fig:Cvx}(c) shows the measured DLIF spectra from the three excited states $\tilde{C},\nu_0$, $\tilde{C},\nu_3$, and $\tilde{C},\nu_6$ of SrOPh. With increasing vibrational energy of the $\tilde{C}$ state, the density of states that is possible to couple also increases. As a result, the $\tilde{C}$ state coefficient $c_C$ in the coupled state is reduced as we increase the excitation energy. This is indeed measured in the DLIF spectra, where the $c_C$ peak intensity drops as the excitation energy increases. The result is summarized in Table \ref{tab:Cvx}. In fact, for $\tilde{C},\nu_6$ the $\tilde{C}$ state fluorescence peak become so weak that it is barely measurable.

A reduction of lifetime is also measured for the $\tilde{C}$ state excited levels with increasing excitation energy. However, we note that the lifetime reduction cannot be explained by $\tilde{C}-\tilde{A}$ coupling, since the primary decay route for the $\tilde{A}$ state component is radiative to the ground state, which share similar or even longer lifetimes compared to the $\tilde{C}$ state. We attribute this to a possible dissociative state in the middle. Based on theoretical calculations, the SrOPh $\rightarrow$ Sr + PhO dissociation energy is 2.70 eV (PBE0/def2tzvpp+gd3), which is above our excitation energy to the $\tilde{C}$ state. It is currently not clear that whether a dissociative state exists at lower energy or not.

\begin{table}[h!]
\centering
\caption{Derivation of coupling strength from vibrationally excited $\tilde{C}$ states of SrOPh}
\label{tab:Cvx}
\begin{tabular}{l|llll}
\hline
\begin{tabular}[c]{@{}l@{}} Vibronic\\state\end{tabular} & \begin{tabular}[c]{@{}l@{}} Energy (cm$^{-1}$)\\ (relative to $\tilde{A},\nu_0$)\end{tabular} & \begin{tabular}[c]{@{}l@{}}Density of \\ states $\rho$ (/cm$^{-1}$)\end{tabular} & \begin{tabular}[c]{@{}l@{}} $I_C/I_{total}$ \\ \end{tabular} & \begin{tabular}[c]{@{}l@{}} $H_{coupling}$ (cm$^{-1}$) \\ \end{tabular} \\ \hline
$\tilde{C},\nu_0$   &  1354     &   16      
&  0.37            &  0.1                  \\
$\tilde{C},\nu_3$  &  1593      &   35
&  0.15            &  0.08                  \\
$\tilde{C},\nu_6$  &  1801       &  73
&  $<$0.05           &  $>$0.075                \\ \hline
\end{tabular}
\end{table}

\clearpage
\section{Franck--Condon factors for higher order vibrational modes}

\begin{figure}[h!]
    \centering
    \includegraphics[width=1\linewidth]{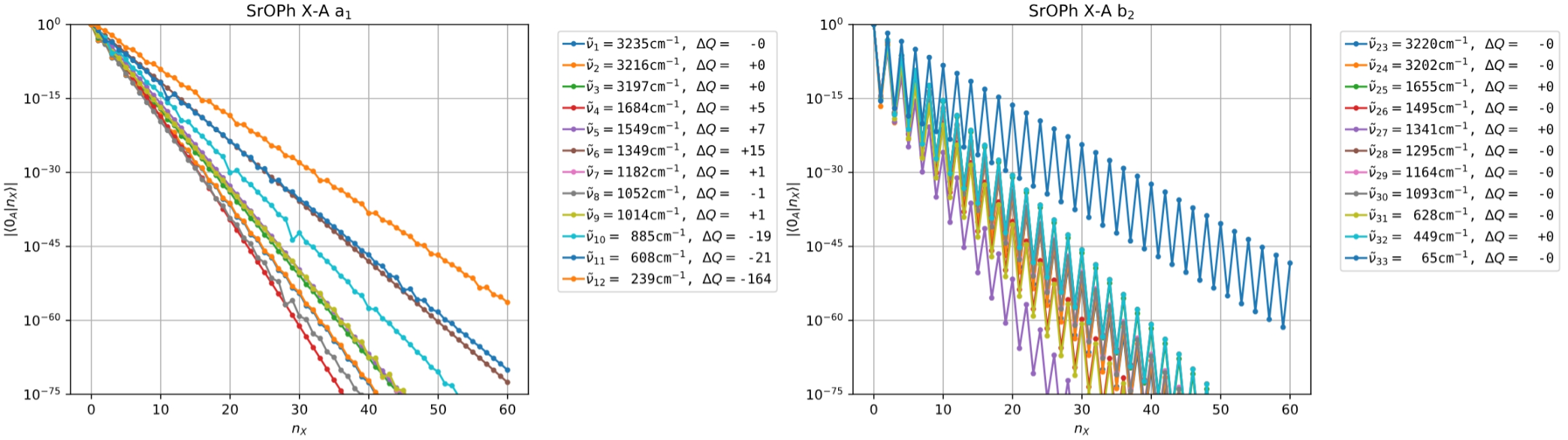}
    \caption{Computed Franck-Condon factors of various vibrational modes of SrOPh, as a function of vibrational quantum number. The plot is grouped by symmetry. The oscillation in the $b_2$-symmetry modes is a result of little change in the harmonic oscillation coordinate ($\Delta Q = 0$). As such, the overlap of wavefunction strongly depends on its odd-even character.}
    \label{fig:S-FCF}
\end{figure}

We estimate the mean coupling strength between the $\tilde{X}$ and $\tilde{A}$ states by scaling the $\tilde{C}-\tilde{A}$ coupling derived in the main text. Compared to $\tilde{C}-\tilde{A}$ energy difference, the $\tilde{X}-\tilde{A}$ energy difference is $\sim15000$ cm $^{-1}$ more, which requires many vibrational quanta to compensate. As such, it is not feasible to compute the transition dipole moments for all these higher order transitions. Instead, we use the Franck-Condon factors to roughly scale the difference. Take SrOPh for example, Fig.~\ref{fig:S-FCF} shows the calculated FCFs for various vibrational modes of SrOPh grouped by their symmetry. The change in FCF differs for different vibrational modes. Given the electronic excitation mainly happens on the metal atom, the overlaps at higher vibrational quanta are generally worse for the vibrational modes that do not involve the metal atom. But on average, we can conclude that the FCF drop by a factor $10^{-1.2}\sim10^{-2}$ per each vibrational quanta. 

In order to estimate an average value of the Franck-Condon factors, we need to know an average number of vibrational quanta that is required to match the $\tilde{A}-\tilde{X}$ energy gap. Here we should note that large density of states is not resulted from a single vibrational mode. Instead, it composes of excitations of many vibrational modes that add up to the desired energy. The mean vibrational quantum number is calculated following the Beyer-Swinehart algorithm~\cite{beyer1973algorithm, stein1973accurate}  used in the MultiWell program~\cite{barker_multiwell_2023, barker_multiplewell_2001, barker_energy_2009}. As such, we determined a mean vibrational quantum number $n_\nu = 45$. Following the FCFs calculated in Fig.~\ref{fig:S-FCF}, we estimated a mean coupling strength $\Omega$ on the order of $10^{-45}$~cm~$^{-1}$. This is well below the limit for laser cooling to be affected by this, as we described in the main text.

We shall note here that the large density of states and the high vibrational quantum number are heavily contributed by the lowest-frequency bending modes ($\nu \simeq60$~cm~$^{-1}$). If these states are excluded, the density of states quickly drops to $10^6$~/cm~$^{-1}$, five orders of magnitude lower. The mean vibrational quantum number also drops to 20. Nonetheless, the condition is still satisfied and the non-adiabatic transition to the $\tilde{X}$ state does not need to be considered.

\newpage
\section{Dependence on molecular size}
In order to verify the dependence of molecular size, we also carried out similar analysis for the SrOPh-diamondoids series that are presented in our recent work.~\cite{lao_bottom-up_2024} These molecules are attached to adamantane and diadamantane structures and have far more atoms compared to SrOPh. As such, they have a much higher density of states at the $\tilde{A}-\tilde{X}$ transition energy. They also have a few low-vibrational frequency modes, which contributes significantly towards this density of states.

The density of states first increases exponentially per each added atom. But the increase gradually slows down, partly due to the fact that the low-frequency modes have a minimum spacing. All these density of states are well within the limit we proposed in the main text. Furthermore, the average quantum number for these complex molecules are much larger compared to SrOPh, which results in the even weaker FCFs and mean coupling strengths. This explains the very favored optical cycling we observed with these molecules.
In general, it appears that non-adiabatic coupling with the ground state $\tilde{X}$ does not need to be concerned.

\begin{table}[h!]
\centering
\caption{The dependence of $\tilde{X}-\tilde{A}$ coupling as a function of molecular size}
\begin{tabular}{l|lllll}
\hline
Molecule    & SrOPh & SrOPh-CH$_3$  & SrOPh-C(CH$_3$)$_3$ & SrOPh-Ad  & SrOPh-1-diAd \\ \hline
\# of atoms   & 13  & 16   & 25   & 37  & 46       \\
$\rho$ (/cm$^{-1}$) & $1\times10^{11}$ &  $1\times10^{13}$   & $5\times10^{17}$   & $5\times10^{20}$   & $1\times10^{23}$   \\
$n_\nu$ & 45 & 81 & 78  & 79 & 90  \\ \hline
\end{tabular}
\end{table}

\clearpage
\section{$\tilde{C}$ state measurements with varying experimental conditions}

\begin{figure}[h!]
    \centering
    \includegraphics[width=0.75\linewidth]{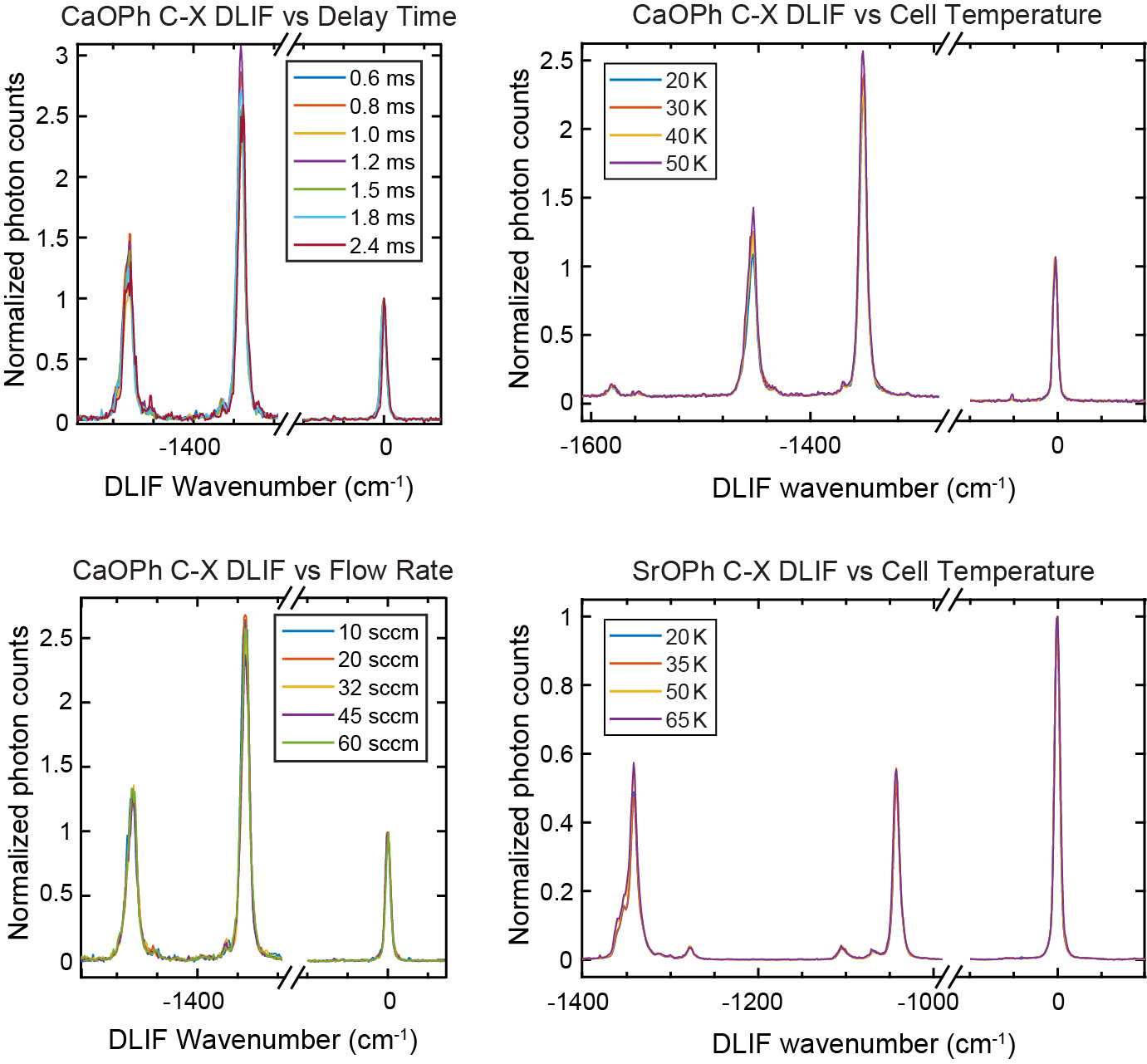}
    \caption{Measured $\tilde{C}$ state DLIFs at different experimental conditions. The relatively ratios between the $\tilde{C}-\tilde{X}$ and $\tilde{A}/\tilde{B}-\tilde{X}$ peak intensities are roughly similar for various buffer gas flow rates, ablation-detection delay times, and cell temperatures, confirming that these peaks are from intrinsic properties of the excited state. In each figure, the spectra are normalized accordingly to their intensities at the zero frequency.}
    \label{fig:C-vary}
\end{figure}

In the DLIF spectra of the $\tilde{C}$ state, we observed additional peaks  at $\tilde{A}\rightarrow\tilde{X}$ and $\tilde{B}\rightarrow\tilde{X}$ transition energy. In order to ensure that these are the intrinsic properties of the excited state, and not resulted from any external factors such as molecular collisions, we performed the same DLIF measurement across various experimental conditions. The varied conditions include chamber temperature, buffer gas flow rate, and detection delay. The results are summarized in Fig.~\ref{fig:C-vary}. For CaOPh, the relative ratio between $\tilde{C}\rightarrow\tilde{X}$ and $\tilde{A}/\tilde{B}\rightarrow\tilde{X}$ peak intensities are around 1:1.2:0.6, with $\sim10\%$ fluctuations. In general, no systematic change of either condition can be confirmed. Similar behaviors are observed in SrOPh measurements with varying temperatures as well. As such, these additional peaks cannot be simply disregarded as some collisional relaxation with the background gas. They are highly likely to decay from the same excitation state, which results from vibronic coupling with the nearby states.

\clearpage
\section{$\tilde{C}$ state measurements in other OCC-functionalized molecules}

\begin{figure}[h!]
    \centering
    \includegraphics[width=1\linewidth]{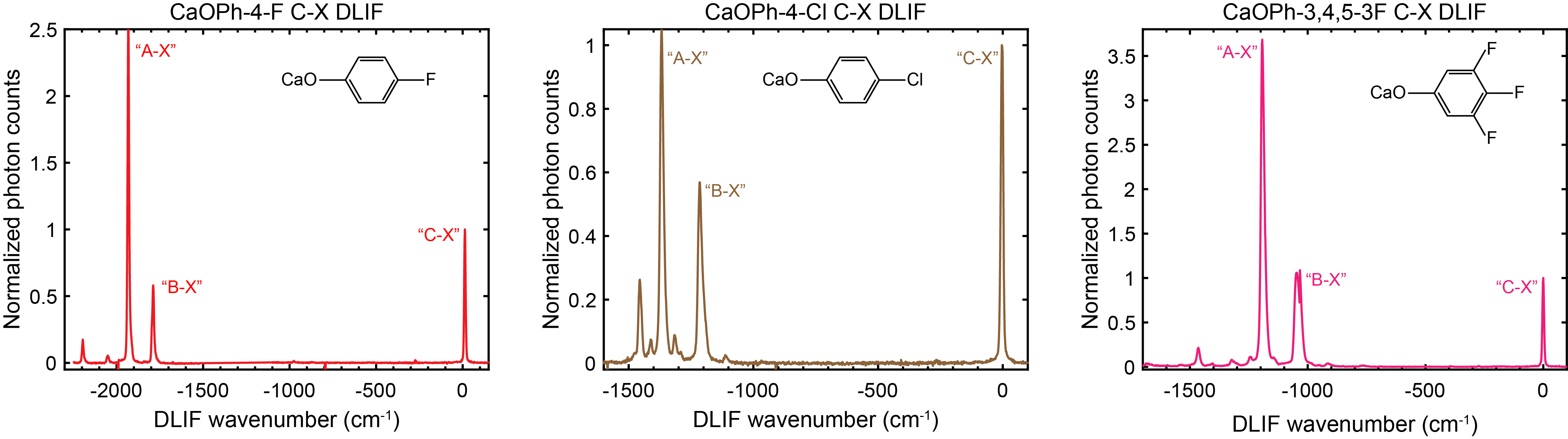}
    \caption{Measured $\tilde{C}$ state DLIFs for some phenol derivatives, CaOPh-4-F, CaOPh-4-Cl, CaOPh-3,4,5-3F. All of these species show similar features in the DLIF.}
    \label{fig:CaOPhX-C}
\end{figure}

We also performed similar spectroscopy with a few substituted phenol species for their $\tilde{C}$ state: CaOPh-4-F, CaOPh-4-Cl, CaOPh-3,4,5-3F. Their DLIF spectra are shown in Fig.~\ref{fig:CaOPhX-C}. In the $\tilde{C}$ DLIF spectra, all these species show peaks at the corresponding vertical relaxation energies of $\tilde{A}\rightarrow\tilde{X}$ and $\tilde{B}\rightarrow\tilde{X}$ transitions, similar to the CaOPh and SrOPh in the main text. These confirms the presence of nonabiabatic coupling between $\tilde{C}-\tilde{A}/\tilde{B}$ states is ubiquitous in many similar molecules.

\begin{figure}[h!]
    \centering
    \includegraphics[width=1\linewidth]{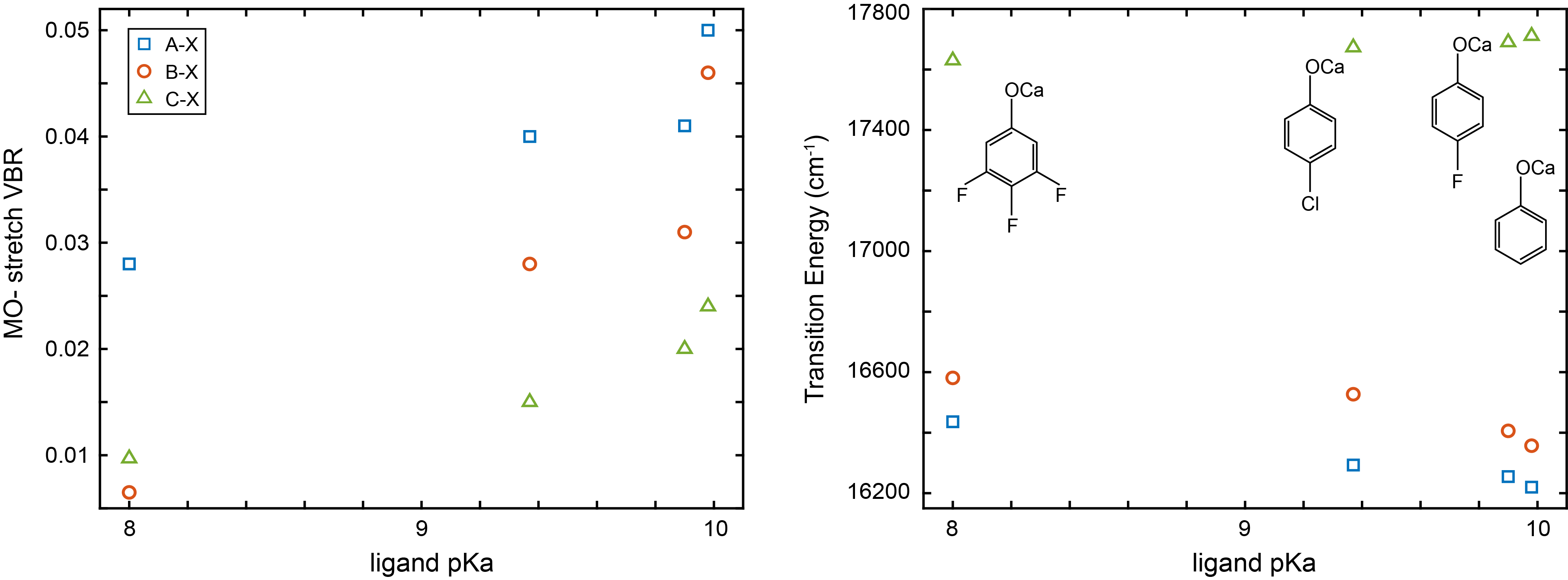}
    \caption{Measured VBR for MO-stretching and transition energies of the $\tilde{C}-\tilde{X}$ transition as a function of pKa for molecules CaOPh, CaOPh-4-F, CaOPh-4-Cl, and CaOPh-3,4,5-3F (Decreasing pKa order). The $\tilde{A}/\tilde{B}-\tilde{X}$ transitions are shown for comparison.}
    \label{fig:CaOPh-X}
\end{figure}

In Fig.~\ref{fig:CaOPh-X} we also plot the transition energy of the $\tilde{C}-\tilde{X}$ transition as well as its VBR for MO-stretching as a function of the pKa of the ligand. Because of the large contributions from the coupled $\tilde{A}$ and $\tilde{B}$ state component, it becomes difficult to obtain a reliable VBR for the 0-0 transition of the $\tilde{C}$ state. Instead, we show the VBR of the MO- stretching mode derived from its peak intensity, which is the most important channel of vibrational-changing decays. Similarly, we notice a better VBR as the acidity (therefore the electron-withdrawing capability) of the ligand is increased, agreeing with previous findings~\cite{dickerson_franck-condon_2021}. The behavior of the $\tilde{A}$ and $\tilde{B}$ states is also shown in the Figure for comparison. Interestingly, the trend for the transition energy is actually reversed compared to the $\tilde{A}$ and $\tilde{B}$ states. With a stronger electron-withdrawing ligand, the transition energy is reduced instead of increased. The trend is visible for the Ca- variety, but less prominent and more irregular for the Sr- variety. 

\end{document}